# The divide between us: Internet access among people with and without disabilities in the post-pandemic era

Edgar Pacheco[a]* and Hannah Burgess[b]

[a]*School of Information Management, Victoria University of Wellington, Wellington, New Zealand;* [b]*Wellington, New Zealand*

The COVID-19 pandemic highlighted the importance of internet access across various aspects of life, from remote work and online education to healthcare services and social connections. As we transition to a post-pandemic era, a pressing need arises to update our understanding of the multifaceted nature of internet access. This study is one of the first attempts to do so. Using survey data from New Zealand adult internet users (n=960), it compares internet connection types, frequency of internet use at home, social media use, and concerns about online risk between people with and without disabilities. Results show people with disabilities have restricted fibre access and higher wireless broadband (a much slower connection type). People with disabilities use social media platforms less and are more concerned about certain online risks. The findings highlight persistent disparities in internet access for people with disabilities in the post-pandemic era. Implications of the study are discussed.

Keywords: internet access; digital divide; disability; broadband divide; digital inequality; digital inclusion, pandemic.

**Introduction**

As the COVID-19 pandemic took hold globally in early 2020 our reliance on internet access accelerated as an essential means for work, education, healthcare, and social connections among other activities (Lai & Widmar, 2021). While the benefits of the internet are vast, people with disabilities have encountered longstanding challenges that result in inequalities when compared to individuals without disabilities (see Dobransky & Hargittai, 2006; Perrin & Atske, 2021; Pettersson et al., 2023). As we transition to the post-pandemic era, it is crucial to gather further evidence that can help address inequalities rather than exacerbate them, and advance equitable internet access for people with disabilities.

In the first year of the COVID-19 pandemic an emerging quantitative literature looked at the pandemic's impact on people with disabilities' internet experiences and related behaviours and attitudes. For instance, scoping research (Jesus et al., 2021) on the health and social impacts of pandemic-related measures (e.g., lockdown and social distancing) found that people with disabilities faced challenges to accessing online health, education, and support services due to poor design of the digital platforms used for these purposes. In the United States, a representative study (Dobransky & Hargittai, 2021) conducted in the first weeks of the pandemic reported that, compared to individuals without disabilities, people with disabilities were more engaged with information about COVID-19. Meanwhile, in South Korea, Cho and Kim (2022) compared the internet usage between adults with and without disabilities in relation to different pandemic-related online services. They found that participants without disabilities reported increased internet use while those with disabilities indicated that their use of the internet remained similar to the pre-pandemic period. Using the same dataset, another study (Park, 2022) found that during the pandemic usage and perceived usefulness of COVID-19-related online services varied according to disability type. For example, participants with hearing impairments reported the lowest usefulness of internet/mobile information services, while people with visual impairments reported the lowest usefulness of internet/mobile delivery services. Meanwhile, internet/mobile subscription services were also rated the lowest by people with hearing impairments. Apps for government subsidies received the lowest rate for usefulness from people with hearing impairments followed by those with language impairments and visual impairments (Park, 2022). A UK-based study with 571 adults with mild to moderate intellectual disabilities (see Caton et al., 2022) found that participants used the internet mainly to connect with family and friends, access social media tools, and share activities online with other people (e.g., playing video games). For these participants, the possibility of forming social connections was considered the most valuable aspect of the internet during the pandemic (Caton et al., 2022). In summary, quantitative inquiry highlights the critical role of the internet for people with disabilities in the beginning of the pandemic; however, this came with challenges affecting their experiences with and attitudes towards the online medium.

Qualitative-based inquiry has also contributed to understanding people with disabilities' internet experiences and perceptions during the pandemic. For instance, in China, Dai and Hu (2022) explored the support actions of members of an online disability support network set up via a social media platform. Members of the network developed content for people experiencing different types of disabilities, filling a gap not addressed effectively by the government. Similarly, Ellis et al. (2021) explored the experiences of a group of blind and low vision participants in Australia. The authors described participants' creative use and adaptation of smartphones and apps to deal with accessibility barriers when trying to use COVID-19-related health emergency services. Finally, in their study of people with intellectual or developmental disabilities, Spassiani et al. (2023) pointed out that attending online events and meetings helped participants to connect with others and mitigate feelings of social isolation in the first year of the pandemic. To sum up, these qualitative studies show that people with disabilities used proactive behaviour and agency to overcome difficulties and meet their own needs during the challenging times of the COVID-19 pandemic.

In general, these studies provide comparative evidence of the online experiences of people with disabilities during COVID-19. However, as we shift into the post-pandemic era, the new landscape requires more evidence to provide clarity on whether inequalities in internet access persist, what they are, and what policy interventions may be appropriate.

*Internet access*

It is well established that internet access is a complex and evolving phenomenon. As van Deursen and van Dijk (2015a) note, internet access is multifaceted, meaning that it needs to be defined and investigated in terms of users' motivations, material access, skills and usage. Motivations refer to either positive or negative attitudes towards the internet and digital technologies, while material access concerns actual, physical access to software, hardware, and internet connection. Another aspect of internet access relates to skills, which pertains to competency to use the internet and other digital tools. Finally, usage refers to the frequency and length of time using the internet (van Deursen & van Dijk, 2015a).

In the early years of the commercialisation of the internet in the 1990s, access was simply defined as having or lacking physical access to software and/or hardware (van Deursen & van Dijk, 2015a). While physical access continues to be a relevant area of inquiry in internet research, now it is recognised that a comprehensive definition of internet access also involves the understanding of skills, motivations, attitudes, and expectations of getting physical access (Dijk, 2017). Some add that the quality of the access, such as broadband speed, and affordability also need to be included in the analysis (McClain et al., 2021; Riddlesden & Singleton, 2014; Vicente & López, 2010). Over the last decade, the multifaceted and changing nature of internet access has not only been embraced by academic fields but also policy circles and governments around the world. This is because internet and related technological innovations have rapidly advanced over the last few decades, moving from simple asynchronous websites to social media, virtual reality, and artificial intelligence. It is also because the binary distinction between the "haves" and "have nots" is insufficient to understand the complexities of internet adoption and use (van Deursen & van Dijk, 2015b). In countries such as New Zealand, the multifaceted understanding of internet access has informed the development of the Government's digital inclusion work programme (see digital.govt.nz 2022).

*Digital divide and disability*

The digital divide is a term coined in the mid-1990s and refers to the evolving and complex disparities in access to the internet as well as other digital tools (Dijk, 2017; Litchfield et al., 2021). The impact of these disparities has also been identified for people with disabilities (Dobransky & Hargittai, 2006; Macdonald & Clayton, 2013). In New Zealand, research evidence in this domain is scarce. However, one study found that 17.2% of people with disabilities do not have access to the internet, a rate significantly higher than that of non-disabled people in the country (4.7%) (Grimes & White, 2019). A government report based on 2013 census data showed that internet access decreased as people with disabilities aged (MBIE & DIA, 2017). Despite efforts for more inclusive legislation and web accessibility standards, inequalities still persist (Scholz et al., 2017). Some evidence suggests that the divide is being bridged (Scanlan, 2022) as an increasing number of people with disabilities are accessing and interacting through digital tools (Perrin & Atske, 2021). However, despite these apparent improvements, people with disabilities still experience inequitable outcomes, restricting their opportunities to access education, work, health services, entertainment and social connections, and ability to participate in broader society through digital tools (Perrin & Atske, 2021).

Evidence also suggests that the digital divide is intersectional, influenced by factors such as level of education, socioeconomic status, place of residence, occupational status, and individual characteristics (Duplaga, 2017; Litchfield et al., 2021). In this sense, a person's social categories and/or group identities can have varied effects on life outcomes such as health or internet use (Roy et al., 2020). When looking at the digital divide in older adults, Liu (2021) found that, among regular internet users, white females with a good socioeconomic status attained good quality of life and low loneliness scores. However, black and minority ethnic females with poorer socioeconomic status, despite regular internet use, had lower quality of life and greater loneliness scores. Meanwhile, another study (Medero et al., 2022) found that older racial and ethnic minorities tended to access the internet on home and public computers less frequently.

In the context of disability, the digital divide is also experienced differently according to the type of impairment and degree of support need (Dobransky & Hargittai, 2006; Duplaga, 2017; Johansson et al., 2021). An age gap in internet use and skills has also been suggested, with young people with disabilities being more active and competent internet users than older users with disabilities (Pacheco et al., 2021).

As Dijk (2017) points out, existing research on the digital divide has concentrated on its causes, with less understanding of the impacts of unequal access and what it means for individuals. Considering that research on the digital divide and disability has been overlooked (Johansson et al., 2021; Scholz et al., 2017), investigating the experiences of internet users with disabilities not only is worthwhile but necessary.

*The internet and disability*

Scholarly work on the internet, and related digital technologies, in terms of disability has been guided by the opposing stances of the medical and social models of disability (Pacheco et al., 2019, 2021). As outlined by the medical model, disability is defined as the consequence of the impairment of bodily functions and structures, which may involve the mind, and can be originated by illnesses, injuries, or health conditions (Haegele & Hodge, 2016). This model has been criticised for placing emphasis on the individual's ability to carry out tasks and prioritising medical interventions aimed at "fixing" or "curing" the individual, without considering the role of social and environmental factors in disability (Haegele & Hodge, 2016). The medical model sees digital technologies from a techno-deterministic standpoint. This perspective emphasises the autonomous and social-shaping characteristics of technology (Dafoe, 2015). According to this perspective, the approach to technology frequently revolves around developing devices that enable individuals with disabilities to overcome their limitations, or function more effectively in society (Roush & Sharby, 2011). The medical model's view on technology asserts that people with disabilities will be ultimately assisted as the 'problems' they face will be compensated or attenuated by technological tools. By placing the cause of disability in the individual, the medical model has had, and still has, a significant influence in the development of assistive technologies (Frauenberger, 2015) as they are developed to improve the functional capabilities of disabled people.

In contrast, the social model holds society to be disabling. According to the social model, society's inability to offer suitable services and adequately guarantee that the needs of disabled people are considered and met is what causes disability, not personal limitations or impairments (Oliver, 1983). In line with this and in the context of technology, proponents of the social model assert that digital tools need to be critically studied because disability is also formed in and through them (Goggin & Newell, 2003). To describe the context of disability and digital technologies, Goggin and Newell (2003) coined the term 'digital disability'. While acknowledging that digital tools, including the internet, can assist people to compensate for their impairments, some proponents of the social model caution that there is a "high price" associated with the perpetuation of borders "between abled and disabled, and normal and deviant" (Moser, 2006). Similarly, it is contended that disparities occur when digital tools are inaccessible to people with disabilities or when obstacles like expensive setup fees, insufficient technical assistance, and exclusive design guidelines hinder their use (Watling, 2011). From a social model perspective, technological tools designed to help can in fact hinder, and perpetuate the social and cultural aspects of exclusion that disabled people already experience (Moser, 2006).

However, while the internet and related digital technologies have brought challenges for people with disabilities, they have also provided opportunities (Chib & Jiang, 2014). Since the inception of the internet, people with disabilities have dealt with issues such as inaccessible design, which make digital tools difficult to use for those experiencing sensory, cognitive, and physical disabilities (Chadwick & Wesson, 2016). Although in recent years an increasing number of people worldwide have been accessing the internet (Poushter, 2016), people with disabilities are still less likely to own a computer or a smartphone (Perrin & Atske, 2021), or be online compared to people without disabilities (Dobransky & Hargittai, 2016), and face a lack of compatibility between some hardware/software and assistive technologies. More recently, the development of Artificial Intelligence (AI) has, for instance, uncovered algorithmic bias against people with disabilities in relation to job recruitment. Despite existing de-biasing measures, pathologisation of people with disabilities is still reinforced (Tilmes, 2022). Yet research shows that, from the experience of people with disabilities, the internet provides a sense of control and empowerment (Chib & Jiang, 2014) and an opportunity to enhance the quality of social interaction, build meaningful connections, and reduce perceptions of loneliness (Chadwick et al., 2013). Tools such as social media also support knowledge sharing and collaboration (Pacheco et al., 2017), self-representation in social media (Haller, 2024), participation and engagement in their communities (Gale & Bolzan, 2016), and the development of self-determination skills (Pacheco et al., 2019). Thus, while the internet and related technologies pose challenges, they also provide opportunities for people with disabilities in their everyday lives.

*The current study*

In light of the above, the purpose of the current study is to expand the understanding of internet access in terms of disability in the post-pandemic era. The study seeks to answer the following question: Are there significant differences between people with and without disabilities regarding aspects of internet access, namely frequency of access at home, type of internet connection, frequency of social media use, perceptions about the benefits of the internet and level of concern about online risks? By exploring the New Zealand context, where little evidence about disability and internet access is available, this study contributes to tracking the current extent of the digital divide in the aftermath of the COVID-19 pandemic.

**Method**

*Design*

Data for the current study came from the *New Zealand's Internet Insights*, an online survey of New Zealanders who are internet users and aged 18 and over. The survey was administered by Kantar Public, a market research company, which employed online consumer panels to recruit participants. The market research company also used a combination of pre-survey quotas to obtain a closer representation of the target population in terms of age, gender, and region. Pre-survey quotas include the number of pre-determined targets or observations required to achieve given criteria and to avoid sampling bias (Lambrecht et al., 2023). Post-survey weighting was also applied based on population estimates from Stats NZ. Post-survey weighting is a statistical technique frequently used after data collection to adjust for nonresponse and other biases in the survey data which helps to improve the representativeness and validity of survey findings (Mercer et al., 2018).

    The use of an online survey for this study was an appropriate methodological choice considering that its use for policy and social research is growing (Lehdonvirta et al., 2021; Sue & Ritter, 2012). More importantly, online surveys have been shown to be a useful tool for exploratory research investigating aspects of people's interaction with and through digital technologies (Pacheco, 2024b). In addition, online surveys are cost-effective and easier to administer (Evans & Mathur, 2018) and the increasing rate of internet access in New Zealand (Pacheco, 2024a) makes it easier to access varied groups of the population as well.

    Data collection was conducted from the 7th to the 14th of November 2022. This was nearly two months after the New Zealand Government ended the COVID-19 Protection Framework. The framework, which ran from December 2021 to September 2022, was set out to help manage life with the Omicron variant. The rules for different 'traffic light' settings were removed because of falling reported COVID-19 cases, a highly vaccinated population, and increased access to antiviral medicines to treat COVID-19. Further description of the framework can be found on a dedicated New Zealand Government website (see New Zealand Government, 2022).

    The maximum margin of error on the total group (n=1,001) is ± 3.1% at the 95% confidence interval.

Note that data for this study was provided to the authors by InternetNZ on request. The organisation made available the anonymised raw survey data and data dictionary to the authors who performed independent secondary data analysis.

*Sample*

Of the participants who undertook the full survey (n=1,001), 960 provided information about their disability status. The final sample for this paper comes from this group, with 166 (17.3%) indicating they have a disability while 794 (82.7%) said they do not have a disability.

Regarding the demographics of participants with disabilities, 60.2% were males and 38.6% were females. Two participants (1.2%) identified as non-binary. Regarding age groups, participants with disabilities in our sample were distributed as follows: participants aged 18–29 and 30–49 years old each represented 21.1%, while those aged 50-64 years old represented 26.5%. Finally, 31.3% was the distribution for those aged 65 years and older. Of participants without disabilities, males represented 47% while 53% were females. In terms of age groups, 21.4% of participants were in the 18–29 age group, and 36.5% were between 30 and 49 years old. Participants aged 50-64 and 65 and older represented 23.7% and 18.4% respectively.

*Ethical considerations*

During the planning and implementation of the data collection stage, the market research company followed the Code of Ethics of the New Zealand Research Association, to which it is a signatory. The market research company obtained and recorded informed consent from all participants. Participants were informed about the purpose of the survey, the topics included, and how information would be used. It was important that a clear, sensitively worded survey introduction was used not only to explain the aim of the study but also to ensure participants were not surprised by the questionnaire content. The survey introduction also informed participants about their right to not take part in the research and right to decline to answer any questions. It also guaranteed participants their right to withdraw from the survey at any time (including during the survey itself). Information about storage of data was also shared with participants. The market research company in charge of data collection ensured that participants' personal information was not associated with their responses; thus confidentiality was assured.

*Measures*

*Disability status*

Disability status was the independent variable. The definition of disability used by Stats NZ's (2014) *Disability Survey 2013* was applied to identify participants who self-reported whether they experience a disability. Participants were asked the following question: 'Do you have a long-term disability or impairment (lasting six months or more) that makes it more difficult for you to do everyday tasks, that other people find easy?' The options listed to answer this question were: 'Yes', 'No', 'Don't know', and 'Prefer not to say'. Participants who chose as an answer 'Don't know' or 'Prefer not to say' were excluded from analysis (n=41).

*Internet access*

The survey included questions aimed at gathering evidence about participants' experiences and attitudes with different aspects of internet access. One of the questions looked at how often participants access the internet at home. For this question the following options were included: 'Once a day or more often', 'Two or three times a week', 'Once a week', 'Two or three times a month', 'Once a month', 'Less than once a month', and 'Never'.

The survey also asked participants about the type of internet connection they have at home. The response options for this question included the following: 'Fibre (any connection)', 'ADSL', 'VDSL', 'Wireless broadband', 'Satellite', 'Don't know', and 'None'.

To measure the frequency of participants' use of specific social media platforms, the survey included the following question: 'How often do you use the following social media channels or messaging services?' Then participants were specifically asked about these social media tools: Facebook, Instagram, Twitter, LinkedIn, WhatsApp, and TikTok. The options listed for each social media tool were: 'Once a day or more often', 'Two or three times a week', 'Once a week', 'Two or three times a month', 'Once a month', 'Less than once a month', and 'Never'.

To explore participants' attitudes towards the internet the following question was asked: 'There are positives and negatives to the internet, but overall do you think the positives outweigh the negatives?'. The response options included 'Yes', 'No', and 'I don't know'.

Then, a multiple-choice question was asked: 'Which of the below do you think are the key benefits of the internet?' For this question participants were asked to think about the key social benefits, regardless of whether they personally use the internet for this reason(s). The list of responses included 16 items. For this question, only descriptive analysis was conducted.

Finally, participants were asked about how concerned they were about different online risks (i.e., cyberbullying, online extremism, misleading or wrong information, conspiracy theories, and sharing of discriminatory content). Conceptually, online risks do not necessarily cause harm but when harm happens it has negative impacts on people's wellbeing (Livingstone, 2013). For each of the five risks included in the current study, participants were asked to think about society as a whole rather than anything they may or may not have personally experienced. To this end, participants could rank each of the five online risks via Likert-type items which included the following scale: 'Extremely concerned', 'Very concerned', 'A little bit concerned', 'Not very concerned', 'Not at all concerned', and 'Don't know'. Participants who selected 'Don't know' were excluded from analysis.

*Data analysis*

All statistical analyses were conducted with Jamovi software, version 2.3 (The jamovi project, 2023). All variables were categorical. First, descriptive statistics such as frequency distributions and percentages were obtained to summarise the data. Then, we applied bivariate inferential statistics. Specifically, the Chi-square test of independence was performed to investigate whether there were statistically significant associations between two categorical variables. The Chi-square test of independence solely assesses associations rather than causal relationships; therefore, it does not offer any insights or conclusion regarding causation. The alpha level was set at 0.05. Additional Cramer's V analysis was conducted to observe the strength of association between variables. We used Rea and Parker's (2014) interpretation to observe the strength of association: a Cramer's V below 0.10 means a negligible association, between 0.10 and below 0.20 indicates a weak association, between 0.20 and below 0.40 represents a moderate association, and between 0.40 and 0.60 indicates a relatively strong association.

Note that, to support the results described in this article, supplementary online material is provided. The material includes the contingency tables with the number of observations in each combination of groups performed on the data. This material can be accessed via the following link: https://figshare.com/s/007e7f3cc2f5cb19866f

**Results**

*Frequency of internet access*

Participants were asked how often they use the internet at home. In general, most (95.1%) reported that they did once a day or more. A small 2.5% did so two or three times a week. When analysing the findings based on disability status, there were statistically significant differences. People without disabilities (96.5%) showed a higher tendency to utilise the internet daily from home compared to those with disabilities (88.6%).

*Broadband connectivity at home*

Overall, 66.9% of participants said they have fibre connection at home while 20.1% connect via wireless broadband. Asymmetric digital subscriber line (ADSL) was the third most common type of internet connection at home (5.0%) among all participants.

The results also show a significant association between disability status and the type of internet connection participants have in their home. Participants with disabilities were less likely to have fibre at home (55.3%) than those participants without a disability (69.3%). In contrast, participants with disabilities were more likely to connect via wireless broadband connection (26.4%) compared to participants without disabilities (18.8%).

*Social media use*

The study also explored the frequency of use of six specific social media tools. When looking at disability status, the Chi-square test of independence only found significant associations with Facebook, Instagram, and WhatsApp but not Twitter, LinkedIn, and TikTok. Regarding Facebook, it was found to be the most popular social media tool for both participants with and without disabilities, but there were nearly 10 percentage points difference between these two groups. Participants without disabilities reported a higher rate of frequent use of Facebook daily (63.7%) compared to participants with disabilities (53.6%). Similarly, the finding regarding the frequency of Instagram use was significant. In this respect, 37.9% of participants without disabilities used

Instagram once a day or more often. This rate was about 16 percentage points higher than the rate reported by participants with disabilities (21.7%). Meanwhile, the use of WhatsApp once a day or more often was higher among participants without disabilities (25.6%) than participants with disabilities (16.9%).

Frequency distributions and percentages for no statistically significant results (i.e., Twitter, LinkedIn, and TikTok) can be found in the supplementary online material.

*Attitudes towards the internet*

The survey asked participants if they believe the benefits of the internet outweigh the drawbacks. The majority of participants (92.7%) replied positively. However, there were no statistically significant differences in the rates reported by participants with and without disabilities.

The study also included a multiple-choice question regarding participants' perceptions of the internet's key benefits. The biggest benefits of the internet for people with disabilities were access to information (71%), and online shopping (69%). Other benefits included ease of communication with friends and family (66%), and ease of access to goods and services (57%). Meanwhile, those without disabilities identified the most important benefit as ease of communication with friends and family (83%). They also identified access to information (81%), online shopping (72%), and working from home (65%) as beneficial.

*Attitudes towards online risks*

The association of disability status and concern about five online risks was tested (i.e., cyberbullying, online extremism, misleading information, conspiracy theories, and hate speech).

The results reveal a significant association between disability status and the level of concern about conspiracy theories. Participants with disabilities expressed more concern about conspiracy theories compared to those without disabilities. Specifically, 35.0% expressed being extremely concerned, while 26.3% stated they were very concerned. Conversely, individuals without disabilities reported rates of 23.8% and 24.8% for being extremely concerned and very concerned, respectively.

A significant association was also found between disability status and concern related to hate speech. Participants with disabilities revealed high levels of concern, with 44.4% expressing extreme concern and 23.8% indicating they were very concerned about hate speech. In contrast, the percentages among non-disabled participants were 30.1% and 33.0% for extremely concern and very concerned, respectively.

It should be noted that there were no statistically significant differences in disability status and the level of concern about cyberbullying, online extremism, and misleading information. The frequency distributions and percentages for these findings are provided in the supplementary online material.

**Discussion**

Having considered the inequitable outcomes people with disabilities are experiencing in New Zealand in the wake of COVID-19, this study also provides a stocktake of trends in internet access and, and more importantly, extends research on disability and the digital divide after the pandemic in the New Zealand context. The findings are further discussed below.

*A disability broadband divide*

One of the key findings of this study regards the disparities in broadband connection. Research shows that, overall, most New Zealand homes connect to the internet through fibre (Diaz Andrade et al., 2021; Pacheco, 2024a). The current study not only expands this evidence in terms of disability status but also revealed a gap in the quality of access. In our sample, 55.3% of people with disabilities said they connect online via fibre. This was about 14 percentage points lower than that of people without disabilities (69.3%). In contrast, the rate of wireless broadband connection at home was higher among people with disabilities compared to those without disabilities. Similar to Dobransky and Hargittai (2016), people with disabilities are less likely to have access to faster broadband connection than people without disabilities; however, this contradicts Perrin and Atske (2021) who found no significant differences between these two groups in terms of high-speed broadband access at home. What is more, as the current study uncovers, disabled people tend to connect online via wireless broadband, a type of connection with lower download/upload speeds and more frequent dropouts than fibre, and with the highest latency of all technologies apart from satellite, including slower speeds than VDSL (Commerce Commission, 2022).

The finding highlighted above suggests the existence of a disability broadband divide which not only refers to access to broadband connection but also disparities in the quality of high-speed internet access. As noted in the Introduction section of this article, within the internet research community, there is a call for including quality of broadband speed in the study of internet access (see Riddlesden & Singleton, 2014; Vicente & López, 2010). Our findings offer empirical evidence that supports the necessity for further exploration in this area and, additionally, identify a connectivity disparity affecting people with disabilities.

Furthermore, existing evidence about the broadband divide has centred on the inequalities experienced by people living in rural areas (Riddlesden & Singleton, 2014) and more recently those from older age groups (Pacheco, 2024a). Our findings expand this evidence by showing that disparities in high-speed broadband also affect people with disabilities. As discussed in the Introduction section of this article, access to the internet has been important for people with disabilities especially during the COVID-19 pandemic. High-speed broadband is critical to enhance access to healthcare and government programmes and services as well as resources such as education, jobs, communication, and information which are crucial for people with disabilities. Therefore, to promote inclusion, and enhance the overall quality of life for people with disabilities, supporting access to high-speed broadband is essential. Policy interventions targeting digital inclusion for disabled communities need to focus on ensuring people with disabilities have access to the highest quality fibre access, especially if an unexpected event occurs and requires lockdowns and other restrictions.

Evidence shows that home is where people most often access the internet (Grimes & White, 2019; McClain et al., 2021; Pacheco, 2024a; Scholz et al., 2017). When looking at disability status, the present study also uncovers some differences in this respect in the New Zealand context. Our study showed that daily access to the internet at home is lower among people with disabilities (88.6%) than those with no disabilities (96.5%). Studies conducted before and during the COVID-19 pandemic (see

Duplaga 2017; Cho and Kim 2022) have found similar results. Our study complements this evidence by providing post-pandemic insights. While disability status is related to internet use, other key factors such as the socioeconomic position of people with disabilities has been shown to play a role in access and affordability of the internet and related technologies (Scholz et al., 2017; Vicente & López, 2010). Around the world different approaches have been implemented to support access to the internet for vulnerable populations. In New Zealand, an online service has been put in place to allow access to government services and information without having to pay for the data (i.e., Zero Data), as has a programme offering low-cost pre-paid internet (i.e., Skinny Jump). While these responses make a contribution to better outcomes, it is clear that a more comprehensive approach is needed to bridge the divide between people with and without disabilities' internet access at home.

*Perceptions of the internet*

Our results show that most participants, 9 in 10, believe that the positives of the internet outweigh the negatives. This figure does not vary statistically between people with and without disabilities. However, compared to those without disabilities, people with disabilities reported lower rates of perceived benefits of the internet. For instance, benefits listed in the survey questionnaire such as ease of communication with friends and family, access to information, online shopping and work from home were considered relevant for both groups, but to a lesser extent for those participants with disabilities. This finding suggests that people with disabilities' positive views of the internet contrast with a persistent pattern of inequitable access reported here and in other studies. As digital inclusion remains compromised for disabled communities, policy solutions should not further burden those experiencing inequitable outcomes, and instead focus interventions within the systems that are creating and reproducing inequities in internet access.

*Differences in social media use*

Research evidence on social media use and disability is limited (Ellis & Kent, 2017). Available studies have considered the relationship of people with specific impairments to social media platforms (Caton & Chapman, 2016; Gkatzola & Papadopoulos, 2023; White & Forrester-Jones, 2020). This study provides quantitative evidence on prevalence of use of specific social media platforms amongst people with disabilities and those without. Significant differences were identified as noted above.

While the factors explaining these differences were not explored in the current study, there is a long-standing discussion and abundant evidence about accessibility barriers preventing people with disabilities from fully using and benefiting from these platforms (Ellis & Kent, 2017; Gleason et al., 2020). This may shine light on the differing rates in social media use between people with and without disabilities in our study. Social media plays an important role for people with disabilities in terms of building connections and feeling connected with family and friends as well as others within the disability community (Chadwick et al., 2013; Chen & Li, 2017; Pacheco et al., 2017). Social media tools also provide an important platform for both personal and collective advocacy and control over disclosing one's disability (Kent, 2019). However, opportunities for participation, including building and maintaining social capital (Chen & Li, 2017), could be missed if barriers such as accessibility are not addressed. Further investigation needs to occur on context, activity type and content engaged with. Incorporating inclusive social media design, which accounts for people with disabilities' various needs and abilities (Olbrich et al., 2015) is an obvious first step.

*Concerns about online risks*

This study showed statistically significant differences in the level of concern about hate speech and conspiracy theories as is noted above. Similar to British and American research (see Emerson & Roulstone, 2014; Sherry, 2012), New Zealand-based evidence shows that not only is hate speech victimisation higher among people with disabilities but also that rates have recently increased for them (see Pacheco and Melhuish 2018; 2019). This evidence might explain, to some extent, the higher rate of concern reported by participants with disabilities in the present study.

When looking at the findings on attitudes towards conspiracy theories, a significant association with disability was found. People with disabilities exhibited a higher likelihood of expressing extreme concern (35.0%) regarding conspiracy theories disseminated on the internet compared to their counterparts without disabilities (23.8%). Extensive scholarly work has been conducted on conspiracy theories in recent years, particularly those related to health, following the outbreak of the COVID-19 pandemic. However, a clear gap still exists in the literature concerning the beliefs in conspiracy theories and their impacts on individuals with disabilities. Subsequent research initiatives should work towards designing or adjusting comprehensive measures to further understand the relationship between online conspiracy theory exposure and disability as uncovered in the current study.

Our study showed that disability status was not significantly associated with concerns regarding cyberbullying, online extremism, and misleading information. Future studies could investigate whether other variables have an impact on people with disabilities' level of concern about online risks, and their self-reported experiences of risks and perceptions of harm. In this respect, a fruitful area for further work will be investigating the incidence of online victimisation among adult internet users with disabilities.

**Conclusion**

The findings of the present paper have important implications not only for the understanding of the way people with disabilities use and experience the internet but also for developing interventions aimed at supporting digital inclusion. As shown, despite high rates of internet access, people with disabilities are still disadvantaged by the type and quality of broadband connection. Efforts to meet web accessibility standards, enhance accessible design, and upskill disabled internet users need to be complemented with measures that support better and faster broadband access, including its affordability. In doing so, people with disabilities will have enhanced opportunities to access government information, health services, banking, connecting with people, shopping, and entertainment, especially in challenging times or events such as the recent pandemic. Policy interventions therefore need to be twofold, targeting both access to faster physical infrastructure and developing personal skills and motivations.

Internet access is a human right (Reglitz, 2020) in that that all people must be able to access the internet in order to exercise and enjoy their rights fully. Furthermore, the United Nation's Convention on the Rights of Persons with Disabilities (CRPD) of which New Zealand is a member, include the principles of equality of opportunity and accessibility. As the internet and related technologies have become vital tools for enabling full participation in society, governments therefore have an obligation to ensure equitable internet access.

As previously mentioned, not all disability types face the same set of challenges regarding internet access (Park, 2022). Research framed from the inception to investigate strengths-based insights from those disability communities that have high or better levels of digital inclusion and equality could be useful in developing policy interventions and solutions for those communities that have lower levels of inclusion and equality. Lessons from one disability community may be transferable or adaptable to others and allow for targeted supports and policy interventions. Therefore, further research should investigate the nuances of the varied internet experiences and behaviours from within disability communities themselves.

Being and feeling safe online is essential to enjoy the benefits of digital access and inclusion. Digital opportunities (e.g., communication and social interaction) are mitigated if harm caused by online threats such as discrimination or bullying occurs. It is therefore essential that any policy intervention designed to enhance internet access for people with disabilities is complemented with a strategy and actions to support online safety awareness and skills.

Finally, this paper is a stocktake of current trends in internet access in New Zealand. A natural progression would be to further explore the extent and nature of internet connectivity using a smartphone as, it may help to explain the lower rate of high-speed internet connection at home for disabled people (Anderson, 2019). In addition, gathering longitudinal data is recommended to understand whether trends of internet access have changed for people with disabilities.

Despite its contribution, the present work also faces some limitations. The survey used in the present work did not gather evidence based on the type of disability, or number of disabilities participants identified with. This would be useful to investigate in future work. The findings in this paper are of course influenced by the original survey design and data provided to the researchers. Furthermore, the data were collected at one point in time, meaning the findings represent a snapshot of participants' experiences. To understand whether trends of internet access post-pandemic change over time, longitudinal evidence is needed. Finally, as the study focuses on internet users, future research may also incorporate the experiences of those who, for various reasons, do not access the online medium.

**About the authors**

Edgar Pacheco holds a PhD in Information Systems. Dr Pacheco is Adjunct Research Fellow at Victoria University of Wellington's School of Information Management, and Senior Research Scientist (Social Sciences) at WSP New Zealand.

Hannah Burgess holds a PhD in History from Otago University. Dr Burgess is a Principal Policy Advisor for the New Zealand Public Service.

**Acknowledgments**

The authors would like to thank InternetNZ (https://internetnz.nz/) for providing open access to the data used in this paper.

# Supplementary online material

# The divide between us: Internet access among people with and without disabilities in the post-pandemic era

Available at **https://doi.org/10.6084/m9.figshare.26420485.v1**

**Supplementary Table 1. Frequency of internet access at home by disability status**

| Frequency of internet connection at home | Disability status | | | | | | $p$ |
|---|---|---|---|---|---|---|---|
| | With disability | | Without disability | | Total | | |
| | n | % | n | % | n | % | |
| Once a day or more often | 147 | 88.6 | 766 | 96.5 | 913 | 95.1 | < .001 |
| Two or three times a week | 6 | 3.6 | 18 | 2.3 | 24 | 2.5 | |
| Once a week | 3 | 1.8 | 2 | 0.3 | 5 | 0.5 | |
| Two or three times a month | 6 | 3.6 | 5 | 0.6 | 11 | 1.1 | |
| Once a month | 2 | 1.2 | 1 | 0.1 | 3 | 0.3 | |
| Less than once a month | 0 | 0.0 | 1 | 0.1 | 1 | 0.1 | |
| Never | 2 | 1.2 | 1 | 0.1 | 1 | 0.3 | |
| Total | 166 | 100.0 | 794 | 100.0 | 960 | 100.0 | |

Note: $p$ value was calculated using the Chi-square test of independence.

**Supplementary Table 2. Type of internet connection at home by disability status**

|  | Disability status | | | | | | |
|---|---|---|---|---|---|---|---|
| Type of internet connection at home | With disability | | Without disability | | Total | | p |
|  | n | % | n | % | n | % |  |
| Fibre | 88 | 55.3 | 537 | 69.3 | 625 | 66.9 | < .001 |
| ADSL | 12 | 7.5 | 35 | 4.5 | 47 | 5.0 |  |
| VDSL | 3 | 1.9 | 32 | 4.1 | 35 | 3.7 |  |
| Wireless broadband | 42 | 26.4 | 146 | 18.8 | 188 | 20.1 |  |
| Satellite | 4 | 2.5 | 14 | 1.8 | 18 | 1.9 |  |
| None | 10 | 6.3 | 11 | 1.4 | 21 | 2.2 |  |
| Total | 159 | 100.0 | 775 | 100.0 | 934 | 100.0 |  |

Note: *p* value was calculated using the Chi-square test of independence.

**Supplementary Table 3. Frequency of social media use by disability status – part 1**

| Social media platform | With disability | | Without disability | | Total | | p |
|---|---|---|---|---|---|---|---|
| | n | % | n | % | n | % | |
| Facebook | | | | | | | |
| Once a day or more often | 89 | 53.6 | 506 | 63.7 | 595 | 62.0 | .001 |
| Two or three times a week | 14 | 8.4 | 99 | 12.5 | 113 | 11.8 | |
| Once a week | 15 | 9.0 | 29 | 3.7 | 44 | 4.6 | |
| Two or three times a month | 8 | 4.8 | 16 | 2.0 | 24 | 2.5 | |
| Once a month | 6 | 3.6 | 12 | 1.5 | 18 | 1.9 | |
| Less than once a month | 7 | 4.2 | 31 | 3.9 | 38 | 4.0 | |
| Never | 27 | 16.3 | 101 | 12.7 | 128 | 13.3 | |
| Total | 166 | 100.0 | 794 | 100.0 | 960 | 100.0 | |
| Instagram | | | | | | | |
| Once a day or more often | 36 | 21.7 | 301 | 37.9 | 337 | 35.1 | < .001 |
| Two or three times a week | 15 | 9.0 | 69 | 8.7 | 84 | 8.8 | |
| Once a week | 9 | 5.4 | 36 | 4.5 | 45 | 4.7 | |
| Two or three times a month | 6 | 3.6 | 19 | 2.4 | 25 | 2.6 | |
| Once a month | 8 | 4.8 | 27 | 3.4 | 35 | 3.6 | |
| Less than once a month | 9 | 5.4 | 68 | 8.6 | 77 | 8.0 | |
| Never | 83 | 50.0 | 274 | 34.5 | 357 | 37.2 | |
| Total | 166 | 100.0 | 794 | 100.0 | 960 | 100.0 | |
| Twitter | | | | | | | |
| Once a day or more often | 20 | 12.0 | 58 | 7.3 | 78 | 8.1 | .435 |
| Two or three times a week | 11 | 6.6 | 46 | 5.8 | 57 | 5.9 | |
| Once a week | 6 | 3.6 | 28 | 3.5 | 34 | 3.5 | |
| Two or three times a month | 7 | 4.2 | 28 | 3.5 | 35 | 3.6 | |
| Once a month | 5 | 3.0 | 25 | 3.1 | 30 | 3.1 | |
| Less than once a month | 14 | 8.4 | 96 | 12.1 | 110 | 11.5 | |
| Never | 103 | 62.0 | 513 | 64.6 | 616 | 64.2 | |
| Total | 166 | 100.0 | 794 | 100.0 | 960 | 100.0 | |

Note: *p* values were calculated using the Chi-square test of independence.

**Supplementary Table 4. Frequency of social media use by disability status – part 2**

|  | Disability status | | | | | | |
|---|---|---|---|---|---|---|---|
| Social media platform | With disability | | Without disability | | Total | | *p* |
|  | n | % | n | % | n | % |  |
| LinkedIn |  |  |  |  |  |  |  |
| Once a day or more often | 13 | 7.8 | 71 | 8.9 | 84 | 8.8 | .476 |
| Two or three times a week | 11 | 6.6 | 77 | 9.7 | 88 | 9.2 |  |
| Once a week | 9 | 5.4 | 66 | 8.3 | 75 | 7.8 |  |
| Two or three times a month | 8 | 4.8 | 41 | 5.2 | 49 | 5.1 |  |
| Once a month | 8 | 4.8 | 50 | 6.3 | 58 | 6.0 |  |
| Less than once a month | 19 | 11.4 | 88 | 11.1 | 107 | 11.1 |  |
| Never | 98 | 59.0 | 401 | 50.5 | 499 | 52.0 |  |
| Total | 166 | 100.0 | 794 | 100.0 | 960 | 100.0 |  |
| WhatsApp |  |  |  |  |  |  |  |
| Once a day or more often | 28 | 16.9 | 203 | 25.6 | 231 | 24.1 | .002 |
| Two or three times a week | 17 | 10.2 | 94 | 11.8 | 111 | 11.6 |  |
| Once a week | 11 | 6.6 | 75 | 9.4 | 86 | 9.0 |  |
| Two or three times a month | 8 | 4.8 | 41 | 5.2 | 49 | 5.1 |  |
| Once a month | 4 | 2.4 | 30 | 3.8 | 34 | 3.5 |  |
| Less than once a month | 10 | 6.0 | 76 | 9.6 | 86 | 9.0 |  |
| Never | 88 | 53.0 | 275 | 34.6 | 363 | 37.8 |  |
| Total | 166 | 100.0 | 794 | 100.0 | 960 | 100.0 |  |
| TikTok |  |  |  |  |  |  |  |
| Once a day or more often | 23 | 13.9 | 113 | 14.2 | 136 | 14.2 | .458 |
| Two or three times a week | 11 | 6.6 | 51 | 6.4 | 62 | 6.5 |  |
| Once a week | 9 | 5.4 | 29 | 3.7 | 38 | 4.0 |  |
| Two or three times a month | 1 | 0.6 | 21 | 2.6 | 22 | 2.3 |  |
| Once a month | 7 | 4.2 | 23 | 2.9 | 30 | 3.1 |  |
| Less than once a month | 9 | 5.4 | 64 | 8.1 | 73 | 7.6 |  |
| Never | 106 | 63.9 | 493 | 62.1 | 599 | 62.4 |  |
| Total | 166 | 100.0 | 794 | 100.0 | 960 | 100.0 |  |

Note: *p* values were calculated using the Chi-square test of independence.

**Supplementary Table 5. If the positives of the internet overweight negatives by disability status**

| Positives overweight negatives | Disability status | | | | | | *p* |
|---|---|---|---|---|---|---|---|
| | With disability | | Without disability | | Total | | |
| | n | % | n | % | n | % | |
| Yes | 137 | 90.7 | 671 | 93.1 | 808 | 92.7 | .317 |
| No | 14 | 9.3 | 50 | 6.9 | 64 | 7.3 | |
| Total | 151 | 100.0 | 721 | 100.0 | 872 | 100.0 | |

Note: *p* value was calculated using the Chi-square test of independence.

**Supplementary Table 6. Level of concern about online risks by disability status**

| Online risk | Disability status | | | | | | p |
|---|---|---|---|---|---|---|---|
| | With disability | | Without disability | | Total | | |
| | n | % | n | % | n | % | |
| Cyberbullying | | | | | | | |
| Extremely concerned | 67 | 41.1 | 261 | 33.2 | 328 | 34.5 | .195 |
| Very concerned | 47 | 28.8 | 266 | 33.8 | 313 | 32.9 | |
| A little concerned | 33 | 20.2 | 184 | 23.4 | 217 | 22.8 | |
| Not very concerned | 8 | 4.9 | 52 | 6.6 | 60 | 6.3 | |
| Not at all concerned | 8 | 4.9 | 24 | 3.0 | 32 | 3.4 | |
| Total | 163 | 100.0 | 787 | 100.0 | 950 | 100.0 | |
| Online extremism | | | | | | | |
| Extremely concerned | 62 | 39.0 | 246 | 31.6 | 308 | 32.8 | .322 |
| Very concerned | 42 | 26.4 | 216 | 27.7 | 258 | 27.5 | |
| A little concerned | 32 | 20.1 | 204 | 26.2 | 236 | 25.2 | |
| Not very concerned | 14 | 8.8 | 77 | 9.9 | 91 | 9.7 | |
| Not at all concerned | 9 | 5.7 | 36 | 4.6 | 45 | 4.8 | |
| Total | 159 | 100.0 | 779 | 100.0 | 938 | 100.0 | |
| Misleading information | | | | | | | |
| Extremely concerned | 55 | 34.0 | 200 | 25.4 | 255 | 26.9 | .226 |
| Very concerned | 51 | 31.5 | 265 | 33.7 | 316 | 33.3 | |
| A little concerned | 40 | 24.7 | 220 | 28.0 | 260 | 27.4 | |
| Not very concerned | 11 | 6.8 | 77 | 9.8 | 88 | 9.3 | |
| Not at all concerned | 5 | 3.1 | 25 | 3.2 | 30 | 3.2 | |
| Total | 162 | 100.0 | 787 | 100.0 | 949 | 100.0 | |
| Conspiracy theories | | | | | | | |
| Extremely concerned | 56 | 35.0 | 185 | 23.8 | 241 | 25.7 | .026 |
| Very concerned | 42 | 26.3 | 193 | 24.8 | 235 | 25.1 | |
| A little concerned | 33 | 20.6 | 208 | 26.7 | 241 | 25.7 | |
| Not very concerned | 20 | 12.5 | 127 | 16.3 | 147 | 15.7 | |
| Not at all concerned | 9 | 5.6 | 65 | 8.4 | 74 | 7.9 | |
| Total | 160 | 100.0 | 778 | 100.0 | 938 | 100.0 | |
| Hate speech | | | | | | | |
| Extremely concerned | 71 | 44.4 | 235 | 30.1 | 306 | 32.5 | .004 |
| Very concerned | 38 | 23.8 | 258 | 33.0 | 296 | 31.4 | |
| A little concerned | 39 | 24.4 | 203 | 26.0 | 242 | 25.7 | |
| Not very concerned | 6 | 3.8 | 62 | 7.9 | 68 | 7.2 | |
| Not at all concerned | 6 | 3.8 | 24 | 3.1 | 30 | 3.2 | |
| Total | 160 | 100.0 | 782 | 100.0 | 942 | 100.0 | |

Note: *p* values were calculated using the Chi-square test of independence.

**Supplementary Figure 1. Perceived key benefits of the internet by disability status**

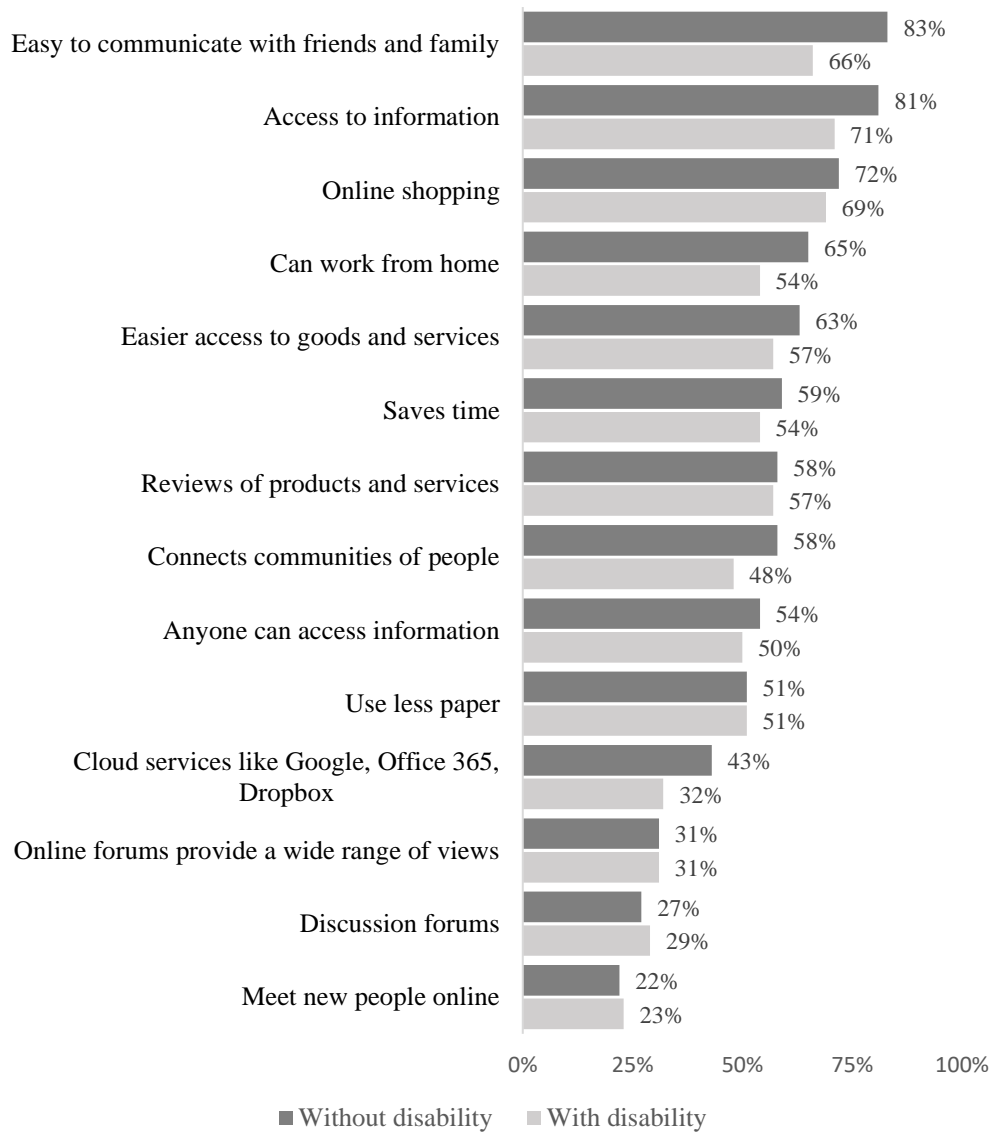

Note: n=960, multiple respond question.